\documentclass[10pt,aps,prl,twocolumn,showpacs,superscriptaddress,nofootinbib,nobibnotes,longbibliography,floatfix]{revtex4-1}

\usepackage[utf8]{inputenc}
\usepackage[T1]{fontenc}
\usepackage{comment}
\usepackage{bm}
\usepackage{mathtools,float,amsmath,amssymb,amsfonts,mathrsfs,eucal,graphicx,tensor,csquotes,accents,commath,chngcntr,siunitx}
\usepackage[dvipsnames]{xcolor}
\usepackage[unicode]{hyperref}
\hypersetup{colorlinks=true, citecolor=MidnightBlue,
            linkcolor=MidnightBlue, urlcolor=MidnightBlue, linktocpage=true}
\usepackage[normalem]{ulem}

\DeclareMathAlphabet{\mathpzc}{OT1}{pzc}{m}{it}
\begin{document}

\title{Non-existence of short hairs for static black holes.}

\author{Soham Acharya}
\email{acharyasoham@iitgn.ac.in }
\affiliation{Indian Institute of Technology, Gandhinagar, Gujarat 382055, India}

\author{Sudipta Sarkar}
\email{sudiptas@iitgn.ac.in }
\affiliation{Indian Institute of Technology, Gandhinagar, Gujarat 382055, India}

\begin{abstract}
The black hole no-short hair theorem establishes a universal lower bound on the extension of hairs outside any 4-dimensional spherically symmetric black hole solutions. We generalise this theorem beyond spherical symmetry, specifically for static, axisymmetric hairy black hole solutions and prove that the ``hairosphere'' must extend beyond the radial extent of the innermost light ring.
\end{abstract}

\maketitle

\section{Introduction}
The no hair conjecture, as originally proclaimed by John Wheeler \cite{Ruffini:1971bza}, suggests that stationary black hole (BH) solutions carry no `hair', i.e., the characteristics of a stationary black hole are entirely determined by conserved charges like mass, angular momentum, and electric charge at the asymptotic infinity.\\

Early works by Bekenstein ruled out specific fields, like scalar, massive vector, and spinor, as matter sources for stationary black hole spacetimes \cite{Bekenstein:1972ky}. These results provided strong support for the black hole "no-hair conjecture". Therefore, the discovery of a colored black hole with Yang-Mills field as hair \cite{Bizon:1990sr} was a significant result that challenged and overturned the celebrated no-hair conjecture. It was soon established that the original version of the no-hair conjecture must be invalid, and there are several hairy BH solutions, such as BHs with skyrmion \cite{Luckock:1986tr}, dilaton \cite{Kanti:1995vq}, and axionic hairs \cite{Campbell:1991rz}.\\

Given the violation of the no-hair conjecture, it is only logical to investigate the essential physical reason that resulted in the existence of these hairy BH solutions. In this context, an important observation from Ref.~\cite{Nunez:1996xv} underscores the fundamental role played by the non-linearity of matter fields. Using Einstein field equations, weak energy condition (WEC), and a non-positive trace condition on matter, it was demonstrated that spherically symmetric black holes cannot possess any short hair. The hairy region, known as the "hairosphere," exhibiting non-linear behavior, must extend at least up to three halves of the horizon radius \cite{Nunez:1996xv}. Intriguingly, this corresponds to the location of the light ring (LR) in $D=4$ dimensions for a Schwarzchild black holes. In \cite{Hod:2011aa} proves that for any hairy black hole in spherically symmetric spacetime, the hairosphere must go beyond the extent of the innermost photon sphere. This "no-short hair" theorem in general relativity (GR) establishes a lower limit for the extent of the hairosphere outside the hairy BH horizon. Therefore, to identify the presence of hair around BHs, it is adequate to investigate the near-LR region exclusively. In other words, since the hairy configuration must extend up to the LR, it is possible to probe the presence of hair in the images of BHs \cite{Cunha:2015yba, Cunha:2019ikd}. Remarkably, the absence of BH short hair transcends beyond general relativity and can be proved without using any gravitational field equations \cite{Ghosh:2023kge}. It is shown that independent of the theory of gravity, the extent of the `hairosphere' must be at least up to the position of the innermost light ring.\\

A limitation of all these results is the explicit use of spherical symmetry. The notion of the light ring and the hairosphere significantly simplify when we consider spherically symmetric black hole spacetimes. Beyond the spherically symmetric case, for a stationary axisymmetric spacetime, the LR is, in general, only a spatially closed null geodesic with a tangent vector field that is always a linear combination of the timelike and azimuthal Killing fields \cite{Cunha:2017qtt}. Then, recent developments regarding the existence of LRs outside the black hole event horizon assert that every stationary black hole solution must have at least one LR outside the event horizon \cite{Ghosh:2021txu}. Therefore, it is natural to ask whether the extent of the non-spherically symmetric BH hair is related to its LRs. We should also note that the original proof of the no-hair theorems is valid beyond spherical symmetry and to a general stationary black hole spacetime. Therefore, it is imperative that we extend the "no-short hair" theorem beyond spherical symmetry. \\

In this work, we aim to explore the static and axisymmetric hairy black hole spacetimes and study the generalisation of the no-short hair theorem. In particular, we prove the following statement: \textit{For static and axisymmetric hairy black hole solutions with certain geometric properties, if we assume weak energy condition and a non-positive trace condition on matter, there will always be angular directions for which the extension of the hair will be beyond the extent of the innermost light ring.} In contrast to spherical symmetry, where the hairosphere extends beyond the light ring in all angular directions, in this scenario, it occurs only within certain angular ranges. We also discuss the implications of our result and possible extensions. 

\section{No black hole short hair theorem: Spherical symmetry}
This section briefly outlines the proof of the black hole no-short hair theorem for spherical symmetry. There have been multiple efforts to understand the properties of the spherically symmetric hairy black hole solutions. The `no-short hair theorem' as proven by \cite{Nunez:1996xv} exploits the spacetime symmetries and the Einstein field equations to prove that the `hair' must extend at least up to a particular distance, and \cite{Hod:2011aa} shows that this particular limit is the radius of the first photon sphere of the spacetime. 
These results start with the metric ansatz for a spherically symmetric metric in the Schwarzchild-like coordinates $(t,r,\theta,\varphi)$ 
\begin{equation}\label{metric_schwarzchild_ansatz}
    ds^2=-e^{-2\delta(r)}\mu(r)dt^2+\frac{dr^2}{\mu(r)}+r^2d\Omega^2 \ .
\end{equation}
Using Einstien's equations and the conservation equation $\nabla_\mu T^\mu_\nu=0$ one has:
\begin{equation}\label{hod_spher_conserv}
    (r^{4}T^r_r)' = \left[(3\mu-1-8\pi r^2 T^r_r)(T^r_r-T^t_t)+2\mu T\right]\frac{r^3}{2\mu} \ .
\end{equation}
Here, $^\prime$ denotes a derivative w.r.t the radial coordinate $r$, and $T$ denotes the trace of the stress-energy tensor. Examining the behavior of the function $r^{4}T^r_r$ in the near horizon region and taking into consideration the regularity on the horizon at $r=r_H$ impose the following boundary conditions at the horizon : 
\begin{equation}\label{hod_boundary_cond_2}
    T^r_r|_{ r_H} \leq 0; \quad \text{and} \quad (T^r_r)'|_{ r_H} < 0 \ .
\end{equation}
Using these boundary conditions at the horizon, it is possible to establish that the function $r^{4}\, T^r_r$ is a non-positive and decreasing function at least up to the radius of the first photon sphere. This implies that the extension of the hair must be equal to or exceed the radius of the first photon sphere. \\

The above proof had a crucial constraint: it is only valid for GR. This proof was generalised by \cite{Ghosh:2023kge}, where the same conclusions were obtained without using gravitational field equations. It started with the ansatz of the following static, spherically symmetric, non-extremal, and asymptotically flat black hole metric:
\begin{equation}\label{rajes_1}
    ds^2=-f(r)\, dt^2+\frac{dr^2}{k(r)}+h(r)\,d\Omega_{(D-2)}^2 \ .
\end{equation} 
The zero of the function $f(r)$ denotes the position of the horizon. Then, owing to the spherical symmetry, we get the following equation: 
\begin{equation}\label{spher_conserv}
    \begin{aligned}
        (h^{D/2} T^r_r)' = 
    \left[\frac{(T^r_r-T^t_t)}{2fh}(fh'-hf')+\frac{h'}{2h}T\right]h^{D/2} \ . 
    \end{aligned}
\end{equation}
As previously, we assume that physical quantities such as $T^r_r$, $T^t_t$, and $T$ remain finite on the black hole horizon at $r=r_H$ (where $f(r_H)=0$). This, along with weak energy condition (WEC) and negative trace energy condition (TEC), dictates that $h^{D/2} T^r_r$ is a non-positive quantity at the horizon and has a local minima located at least at the innermost photon sphere. As a result, the radius of the hair, where the extremum of this function occurs, must necessarily extend beyond that point. This extends the no-short hair theorem for any hairy black hole solution independent of the theory of gravity.\\

The assumption of spherical symmetry plays a crucial role in all these proofs. However, the original no-hair theorems are proven for black hole solutions, which need not be spherically symmetric. Therefore, it's natural to explore the validity of the short hair theorem by relaxing some of the symmetry constraints. In particular, we would like to know if a static, axisymmetric, and asymptotically flat hairy black hole solution can have hairs that are not short, similar to the case for spherical symmetry. Interestingly, as discussed before, the notion of a photon sphere can be extended beyond the spherical symmetry as a light ring, which is spatially closed null geodesic. The existence of such a light ring has been recently extensively explored for both black holes and other compact objects. In the next section, we aim to use these results and find a suitable generalisation of the black hole no-short hair theorem without spherical symmetry.

\section{Beyond spherical symmetry: static and axisymmetric hairy black holes}

We start with a static, asymptotically flat black hole spacetime and consider the decomposition of the spacetime metric as $^{4}g=-V^{2}dt^{2}+ {}^{3}g$. The static domain is characterized by a well-defined positive coordinate $V$, which goes from $V=0$ at the event horizon to $ V \rightarrow 1$ at spatial infinity. The regularity of surfaces $V = \text{constant}$ ensures the quantity $\rho = (dV|dV)^{-1/2}$ is non-vanishing \cite{Heusler}. We introduce the coordinates $\theta$ and $\varphi$ to parameterize the compact two-dimensional $V$ = constant and $t$ = constant surfaces, denoted by $\Omega$, embedded in the three-dimensional Riemannian space $\Sigma$ with metric ${}^{3}g$. We also assume that the spacetime is axisymmetric and possesses a Killing vector $\partial_\varphi$. Then, the metric  is written as:
\begin{equation}\label{metric}
    \begin{aligned}
        ds^2 = -V^{2}dt^{2} + \rho^{2}dV^{2} + g_{\theta\theta}\,d\theta^{2} + g_{\varphi\varphi}\,d\varphi^{2} .
    \end{aligned}
\end{equation}
Where, by construction $\rho, \, g_{\theta\theta}$ and $g_{\varphi\varphi}$ are all functions of $(V,\theta)$. To prevent closed timelike curves, we would also require $g_{\varphi\varphi}>0$. The event horizon at $V = 0$ is considered to be non-extremal, and all physical quantities, including the curvature scalars, are regular at the event horizon. The extrinsic curvature of $\Omega$ embedded in $\Sigma$ is given by,
\begin{equation}
    K_{ab} = \frac{1}{2\rho}\partial_V(g_{ab}) \quad \text{and} \quad a\in(\theta,\varphi) \ .
\end{equation} 
Note, on the horizon, the extrinsic curvature vanishes identically \cite{Heusler}.\\

The asymptotic flatness assumption implies the existence of an asymptotic coordinate system $\{ x^{\alpha}  \}$ in which the metric takes the form, 
\begin{equation}\label{asym_behaviour}
V^2 \rightarrow 1-\frac{2M}{r}+\mathcal{O}(r^{-2}),\,\,  \textrm{as}\,\, r\rightarrow\infty \ .
\end{equation}
Where \(r^2=\delta_{\mu\nu}x^{\mu}x^{\nu}\), and $M$ is associated with the ADM mass of the spacetime, considered to be positive. \\

Let's focus on the $V$ coordinate, where \(V\rightarrow1\) defines the asymptotic limit. Using Eq.~\eqref{asym_behaviour}, we may conclude the following asymptotic expressions of the metric:
\begin{equation}\label{asym_behaviour_2}
g_{\theta\theta}\rightarrow \frac{4M^2}{(1-V^2)^2}, \quad g_{\varphi\varphi}\rightarrow {\frac{4M^2}{(1-V^2)^2}}\text{sin}^2\theta .
\end{equation}
The metric we have been considering, as described in Eq.~\eqref{metric}, represents a solution to a theory with energy-momentum tensor $T_{\mu\nu}$ obeying weak energy condition and trace negativity. 
Given the symmetries of the background solution, the off-diagonal components \(T^\mu_{\varphi}\) must vanish. The independent surviving components include all diagonal terms and off-diagonal terms involving only \((t, V, \theta)\). Notably, the surviving components of the stress tensor are functions solely of \((V, \theta)\). To maintain regularity at the \(V=0\) surface, it is necessary for the scalars $T^\mu_{\nu}T^\nu_{\mu}$ to be finite there.\\

The $V$ component of the energy-momentum conservation equation based on the metric given in Eq.~\eqref{metric} provides us with:
\begin{widetext}
    \begin{equation}\label{conserv_eq}
        \partial_V\left(g_{\theta\theta}^2 T_{V}^{V}\right) = -\left(\left(\frac{2g_{\varphi\varphi}-Vg'_{\varphi\varphi}}{2Vg_{\varphi\varphi}}\right)(T_{V}^{V} - T_{t}^{t}) + \frac{1}{\sqrt{-g}} \partial_{\theta}\left(T_{V}^{\theta}\sqrt{-g}\right) +\rho\left(K^\varphi_\varphi-K^\theta_\theta\right)(3T_{V}^{V} + T_{\theta}^{\theta})-\rho K^\varphi_\varphi T\right)g_{\theta\theta}^2 \ .
    \end{equation}
\end{widetext}
This equation is the generalisation of the Eq.~\eqref{spher_conserv} beyond spherical symmetry. Here, $^\prime$ denotes a derivative w.r.t the coordinate $V$, and $T$ represents the trace of the energy-momentum tensor. As in spherical symmetry, we assume that the components $T^t_t$ and $T^V_V$ decay faster than $(1-V^2)^{4}$ as $V\rightarrow1$. This condition ensures that we are considering a hairy black hole solution, which, in the terminology of \cite{Nunez:1996xv, Coleman:1991ku}, does not have "secondary" hairs. As in case of spherical symmetry, we aim to establish a definite sign of the RHS of this equation by carefully analysing the terms.\\ 

When examining this equation compared to the spherically symmetric scenario, it becomes evident that deviations from spherical symmetry lead to extra terms in Eq.~\eqref{conserv_eq}. Despite this difference, akin to the situation in spherical symmetry, the coefficient of $\left(T^V_V - T^t_t\right)$ will be connected to the extent of the light ring (LR), as we shall see in the discussions to follow.\\

Following the spherically symmetric case, \cite{Nunez:1996xv, Ghosh:2023kge}, to understand the existence of the short hair for this geometry, we first study the near horizon behavior of terms in Eq.~\eqref{conserv_eq}. On the horizon, \(\rho^{-1}\) coincides with the surface gravity \cite{Heusler}, and staticity implies that $\rho$ must be constant at the horizon. We also demand that on the horizon, the curvature scalars be finite. These allow us to construct the following Taylor series expansion of the metric components \cite{Medved:2004ih}: 
\begin{subequations}\label{taylor_expansion}
    \begin{align}
    g_{\theta\theta}(V,\theta) &= g_1(\theta) + \frac{1}{2!}g_2(\theta)V^2 + \mathcal{O}(V^3) \ , \\
    g_{\varphi\varphi}(V,\theta) &= h_1(\theta) + \frac{1}{2!}h_2(\theta)V^2 + \mathcal{O}(V^3) \ ,
\end{align} 
\end{subequations}
using WEC, the finiteness of the RHS of Eq.~\eqref{conserv_eq} at the horizon and taking into consideration the near horizon behavior of the metric components Eq.~\eqref{taylor_expansion}, we see that 
\begin{equation}\label{near_horizon1}
    T^V_V(0,\theta) = T^t_t(0,\theta) \leq 0 \ ,
\end{equation}
which is a generalisation of the first inequality in Eq.~\eqref{hod_boundary_cond_2}. \\

We now assess the on-horizon behavior of the $\theta$-derivative of $T^\theta_V$. To do so, we will use the framework outlined in \cite{Bhattacharyya:2021jhr, Bhattacharyya:2022nqa} to study the behavior of these terms in the near horizon regime. We rely on boost symmetry on the horizon, which emerges from the Killing isometry in the context of stationary black holes, where the Killing horizon remains unchanged when subjected to the transformation induced by this boost \cite{Bhattacharyya:2021jhr}. The boost-symmetry allows us to restrict the form of the field equations of the theory and ensures that,
\begin{equation}\label{near_horizon2}
T^\theta_V\left(0,\theta\right) = \partial_\theta\left(T^\theta_V(0,\theta)\right)=0 \ .
\end{equation}
Using Eq.~\eqref{near_horizon1}, Eq.~\eqref{near_horizon2}, TEC, and the fact that the extrinsic curvature of the co-dimension-2 surface is identically zero at the horizon, we find that at the horizon:
\begin{equation}\label{near_horizon_final}
    \partial_{V}(g_{\theta\theta}^{2}T_{V}^{V})|_{V=0} = 0 \ .
\end{equation}
This is the generalisation of the second inequality in Eq.~\eqref{hod_boundary_cond_2}. We write it as an equality because we are using the $V$ coordinates. It is straightforward to arrive at the second inequality in Eq.~\eqref{hod_boundary_cond_2} using the relationship between $V$ and the radial coordinate $r$ for spherical symmetry.\\ 

Now, we will proceed to evaluate the nature of the RHS of Eq.~\eqref{conserv_eq} away from the horizon. First, we will establish the coefficient of $\left(T^V_V-T^t_t\right)$ as the light ring equation and assign a sign to that term. To achieve this, we start with the definition of a light ring as in \cite{Cunha:2017qtt}: A light ring in a stationary axisymmetric spacetime is a null geodesic such that its tangent vector field can be expressed as a linear combination solely involving the Killing vectors $\partial_t$ and $\partial_\varphi$. This leads to a specific condition on the photon momentum, where both the momentum components, $p_\mu = \dot{p}_\mu = 0$. Here, $\mu$ can take values of $V$ or $\theta$. The conditions for the occurrence of a light ring can be reformulated as a potential $\Psi= \nabla \Psi = 0$ \cite{Guo:2020qwk, Ghosh:2021txu}. From the Lagrangian and using the conserved charges associated with the killing vectors $\partial_{t}$ and $\partial_{\phi}$; $p_{t}=-V^2\dot{t}\equiv E$ and $p_{\varphi}=g_{\varphi\varphi}\dot{\varphi}\equiv\Phi$, we obtain the equation for the potential in the following factorized form:
\begin{equation}
    \Psi = \frac{\Phi^{2}}{V^{2}} \left(\frac{E}{\Phi} - \sqrt{\frac{V^{2}}{g_{\varphi\varphi}}}\right) \left(\frac{E}{\Phi} + \sqrt{\frac{V^{2}}{g_{\varphi\varphi}}}\right) \ .
\end{equation}
Then, for a light ring, the following equations have to be simultaneously satisfied: 
\begin{subequations}
    \begin{align}
        \text{The radial equation:} \quad \frac{1}{V} - \frac{g'_{\varphi\varphi}}{2g_{\varphi\varphi}} &= 0 \label{eq:radial} \ , \\
        \text{The angular equation:} \quad \partial_\theta{g_{\varphi\varphi}}  &= 0 \ . \label{eq:angular}
    \end{align}
\end{subequations}
Note that the radial equation is the same as the coefficient of $\left(T^V_V - T^t_t\right)$ in Eq.~\eqref{conserv_eq}. \\

For the sake of clarity, we can rewrite the radial equation in terms of two auxiliary functions: $L(V,\theta) = 2g_{\varphi\varphi}$ and $R(V,\theta)= Vg'_{\varphi\varphi}$ so that we can write Eq.~\eqref{eq:radial}, as $\Delta(V,\theta)=L(V,\theta)-R(V,\theta)=0$, \cite{Ghosh:2023kge}. In order to study the behavior of the term $\Delta(V,\theta)$ in the interval $0\leq V < 1$, we can consider the following :
\begin{enumerate}
    \item At the horizon, $V=0$, thus $L(0,\theta)>0$ and $R(0,\theta)=0$ 
    \item For the asymptotic limit \(V \rightarrow 1\), using Eq.~\eqref{asym_behaviour_2}, 
        \begin{align*}
            L(V, \theta) &= g_{\varphi\varphi}(V, \theta) \rightarrow \frac{\sin^2{\theta}}{(1 - V^2)^2} \ , \\
            R(V, \theta) &= Vg'_{\varphi\varphi}(V, \theta) \rightarrow \frac{V^2 \sin^2{\theta}}{(1 - V^2)^3} \ .
        \end{align*}
\end{enumerate}
Using these two properties, we see that functions $L(V,\theta)$ and $R(V,\theta)$ must have at least one intersection at $(V=V_p)$, where $V_p$ is the solution to the radial equation Eq.~\eqref{eq:radial} of the light ring, in the region $0\leq V <1$ for all $\theta \in [0,\pi]$ \cite{Guo:2020qwk, Ghosh:2021txu}. Thus, the interval $[0,1)$ is divided into an even number of regions. We can ascertain that $L(V,\theta)-R(V,\theta)<0$ in the outermost region, we must have $L(V,\theta)-R(V,\theta)=2 g_{\varphi\varphi}-V g'_{\varphi\varphi}>0$ in the innermost interval $[0, V_p]$. \\

This analysis proves that the first term in Eq.~\eqref{conserv_eq} must be positive in the innermost interval $[0, V_p]$ with $V_p$ being the solution of the radial equation for the light ring, as long as the WEC holds. \\

We then analyze the behavior of the term $\partial_{\theta}(T_{V}^{\theta}\sqrt{-g})/\sqrt{-g}$ away from the horizon and up to the first light ring. Since we are considering an axisymmetric metric with an axis of rotation, we have,
\begin{equation}\label{axissymmetry}
    g_{\varphi\varphi}(V,\theta) = 
    \begin{cases}
        0 & \text{as }\theta\rightarrow 0 \\
        0 & \text{as }\theta\rightarrow \pi \ .
    \end{cases}
\end{equation}
Note that $T^\theta_V$ cannot have a form that goes as $ \backsim (g_{\varphi\varphi})^{-n}$, where $n$ is some positive real number. Such a form would have caused $T^\theta_V$ to diverge at the poles of all $V=$ constant surfaces, including at the horizon. This also implies that $T_{V}^{\theta}\sqrt{-g}$ vanishes at the poles. Therefore, there must exist at least one interval of $\theta$, $[\theta_i,\theta_f]$ where the function is increasing : 
\begin{equation}
    \partial_{\theta}(T_{V}^{\theta}\sqrt{-g}) > 0 \ .
\end{equation}
From the above two arguments, we can draw the following conclusion - 
for $V\in[0,V_p]$ and $\theta\in[\theta_i,\theta_f]$, where $T_{V}^{\theta}\sqrt{-g}$ is an increasing function of $\theta$,  we set the following signs: 
\begin{widetext}
\begin{equation}\label{conserv_new_signed}
    \partial_{V}\left(g_{\theta\theta}^{2}T_{V}^{V}\right) = -\left( \left.\overbrace{\left(\frac{2g_{\varphi\varphi}-Vg'_{\varphi\varphi}}{2Vg_{\varphi\varphi}}\right)(T_{V}^{V}-T_{t}^{t})}^{\text{+ve}} +\overbrace{\frac{1}{\sqrt{-g}}\partial_{\theta}\left(T_{V}^{\theta}\sqrt{-g}\right)}^{\text{+ve}} \right. + \rho\left(K^\varphi_\varphi-K^\theta_\theta\right)(3T_{V}^{V} + T_{\theta}^{\theta}) -\rho K^\varphi_\varphi T \right )g_{\theta\theta}^{2} \ .
    \end{equation}
\end{widetext}
We have not yet been able to assign a sign to the last two terms. Evaluating $(3T_{V}^{V} + T_{\theta}^{\theta})$ for known matter fields that permit a hairy black hole solution, we discover that this term can exhibit a wide range of values, making it challenging to assign a specific sign. To address this issue, we impose a geometric condition on the embedding of the surface $\Omega$ in the three space $\Sigma$, which makes one of these terms vanish. We demand that the extrinsic curvature obeys $ K^\varphi_\varphi-K^\theta_\theta = 0 $ for these surfaces. This immediately leads to the following condition,
\begin{equation}
   K_{ab} = \alpha(V,\theta)h_{ab},
\end{equation}
where $\alpha(V,\theta)$ is a proportionality factor, which need not be a constant. Therefore, we need the extrinsic curvature to be proportional to the intrinsic metric. Remarkably, a two-dimensional surface embedded in a three-dimensional Riemannian space that obeys such an embedding condition has been studied extensively in the mathematics literature and goes by the name of \textit{totally umbilical embedding} \cite{Chen, Kobayashi}. A trivial example of such an embedding would be a two-dimensional sphere embedded in a three-dimensional Euclidean space. Notably, a geometric sphere is the only totally umbilical convex surface that can be embedded in three-dimensional Euclidean space \cite{Chen, Kobayashi, Ida:2011jm}. However, fortunately, in our case, the three-dimensional space $\Sigma$ is Riemannian; thus, there are non-spherical configurations of the two-dimensional geometry with totally umbilical embedding. We subsequently proceed with the assumption that we are considering the two-dimensional surfaces $\Omega$ to be totally umbilical, and this helps to get rid of the third term in Eq. \eqref{conserv_new_signed}. Along with that, we will further assume that the surface is convex and therefore $K^\varphi_\varphi > 0$. \\

These assumptions allow us to obtain, at least within a non-empty interval $\theta\in[\theta_i,\theta_f]$ :
\begin{equation}
    \partial_{V}\left(g_{\theta\theta}^{2}T_{V}^{V}\right) < 0
\end{equation}
This implies that for this specific range of $\theta$ and for a matter content that satisfies the energy conditions and fall-off conditions, $\left(g_{\theta\theta}^{2}T_{V}^{V}\right)$ starts as a non-positive quantity at the horizon and decreases at least up to the value of $V=V_p$, which is the solution of the radial equation of the light-ring. Following \cite{Ghosh:2023kge, Hod:2011aa}, we define $V=V_{hair}$ as the extent of the hairosphere, where the quantity $\vert g_{\theta\theta}^{2}T_{V}^{V}\vert$ has a local maximum. This, in turn, implies that there must be a non-empty range of the angular coordinate $\theta$ for which the hairosphere must be beyond the radial extent of the innermost light ring. This is the generalisation of the black hole short hair theorem beyond spherical symmetry.\\

We can, therefore, make the following statement: For hairy black hole solutions endowed with staticity and axisymmetry, assuming the weak energy condition and a non-positive trace condition on matter, there will always be angular directions for which the extension of the hair will be beyond the extent of the innermost light ring. Unlike the case of spherical symmetry, where the hairosphere extends beyond the light ring in all angular directions, here it happens only for a range of angular directions, i.e., only for some specific ranges of $\theta$. \\

There is also a crucial difference with the spherically symmetric case; the extension of the hairosphere is only related to the solution of the radial equation for the light ring. So, we cannot ascertain if the solution for the angular equation of the light ring also lies in the same interval. Thus, there may or may not be an actual light ring in the direction the hair extends. The situation simplifies considerably if we consider the existence of reflection symmetry, about some $\theta = \theta_0$ plane, such that, in that plane $\partial_{\theta}(T^{\theta}_{V}\sqrt{-g})$ and $\partial_{\theta}(g_{\varphi\varphi})$ vanish identically. Therefore, in that case, there exists a light ring on the $\theta = \theta_0$ plane, and the hairosphere extends beyond the light ring on that plane.\\

It is also important to note that the validity of the no-short hair theorem requires that we must make some assumptions about the nature of embedding the two-dimensional surfaces, $\Omega$ in the three space $\Sigma$; essentially, we require that the embedding to be totally umbilical. Then, we see that our assumption about the embedding of the co-dimension two surfaces restricts the metric to the form :  
\begin{equation}
ds^2 = -V^{2}dt^{2} + \rho^{2}\,dV^{2} + g(V,\theta)(d\theta^{2} + h(\theta)d\varphi^{2}).
\end{equation}
Where $g(V,\theta)$ and $h(\theta)$ are arbitrary functions of the arguments. For such a static, axisymmetric hairy black hole solution of the above form, the extension of the hair will be beyond the extent of the innermost light ring, at least for a range of angular directions. If the metric further has reflection symmetry, then there must be an angular plane where the hairosphere extends beyond the light ring on that plane.

\section{Conclusion and Discussions}

The black hole no-short hair theorem shows that the region with a non-trivial structure of the hairy matter field or the "hairosphere" must extend beyond the first photon sphere of the spacetime. This is an important observation which may have important observational implications. However, the assumption of spherical symmetry of spacetime is crucial in proving the theorem, a simplification that may not hold in realistic astrophysical scenarios. Therefore, while spherical symmetry serves as a helpful starting point, it is a natural progression to move towards less restrictive use of symmetry and ask whether the original proposition that black hole hairs must extend beyond the first photon sphere still holds. \\

Without using any gravitational field equation, we establish that, for four-dimensional static and axisymmetric hairy black hole solutions, the hair must grow beyond the radial extent of the first photon sphere for some specific ranges of polar angles, as discussed in the previous sections. We have yet to ascertain whether a photon sphere actually exists in these angular directions. Nevertheless, the existence of a reflection symmetry about some plane $\theta = \theta_0$ ensures that on the same plane, the extent of the hair must be beyond the first photon sphere, akin to the original propositions in the case of spherical symmetry. This is of significance to the observation of the black hole shadow. As the hairosphere is not confined to the near-horizon regime, studying the region near the photon sphere exclusively could provide valuable insights into the characteristics of black hole hairs. \\

We further note that the assumption about the two-dimensional surface $\Omega$ being totally umbilically embedded in the three-dimensional space $\Sigma$ is crucial to the validity of our result. Relaxing this constraint on the geometry of the spacetime may be possible by imposing restrictions on the behavior of the matter field, which could be the subject matter for further study. \\

Another possible extension of our result could be to relax the condition of asymptotic flatness. The condition $1/V < g'_{\varphi\varphi}/g_{\varphi\varphi}$ at asymptotic infinity is sufficient to show that the first term of Eq.~\eqref{conserv_eq} is positive, which the asymptotically flat forms of the metric component happen to satisfy. Still, one can look into other asymptotic conditions. It will also be interesting to extend our results to higher dimensions or when assumptions of axisymmetry are relaxed. We leave such possible extensions for future attempts.

\section{Acknowledgment}
We express our thanks to Dawood Kothawala for early collaboration, thorough discussion and valuable comments on our work. We are grateful to Rajes Ghosh for engaging in extensive discussions and providing helpful feedback. We acknowledge Aravinda C S and Gururaja Upadhya for their help in clarifying the properties of totally umbilical embedding. We also thank Shahar Hod, Daniel Sudarsky and 
Hernando Quevedo for their comments. The research of SS is supported by the SERB grant CRG/2023/000545.

\end{document}